\documentclass[aps,prl,amsmath,amssymb,reprint,superscriptaddress]{revtex4-1}%

\usepackage{graphicx}
\usepackage{dcolumn}
\usepackage{bm}

\begin{document}

\preprint{AIP/123-QED}

\title{Active subnanometer spectral control of a random laser}

\author{Marco Leonetti}
\affiliation{ISC-CNR, UOS Sapienza, P. A. Moro 2, 00185 - Roma, Italy}
\affiliation{Instituto de Ciencia de Materiales de Madrid (CSIC), Calle Sor Juana In\'es de la Cruz 3, 28049 Madrid Espa\~{n}a.}

\homepage[]{www.luxrerum.org}\email[]{marco.leonetti@roma1.infn.it}

\author{Cefe L\'opez}%
\affiliation{Instituto de Ciencia de Materiales de Madrid (CSIC), Calle Sor Juana In\'es de la Cruz 3, 28049 Madrid Espa\~{n}a.}


\date{\today}

\begin{abstract}
We demonstrate an experimental technique that allows to achieve a robust control on the emission spectrum of a micro random laser and to select individual modes with sub-nanometer resolution. The presented approach relies on an optimization protocol of the spatial profile of the pump beam. Here we demonstrate not only the possibility to increase the emission at a wavelength, but also that we can "isolate" an individual peak suppressing unwanted contributions form other modes.
\end{abstract}


\maketitle

Standard laser sources, allow to generate light with a high degree of coherence, precise directionality, and gaussian beam profile. Thanks to the progress of nonlinear optics, modern devices allow also femtosecond pulse compression\cite{Haus_mode_lock} and wavelength tunability.\cite{Coldren_tunable,tunablebook} These technological achievements always rely on a precise control of light paths in a macroscopic (usually of the order of the meter) cavity for which any imperfection (like material damage, optical impurity, mechanical instability or misalignment) result in a severe decrease in device efficiency: any form of disorder is detrimental. On the other hand it has been recently demonstrated that disorder may be compensated by various approaches involving adaptive optics: in practice, by introducing an appropriate phase delay to each point of a laser wavefront warped by disorder, it is possible to restore it to its pristine state.\cite{mosk2012controlling, popoff2010measuring} Applications range from the possibility to focus through opaque media both in space and in time\cite{Vellekoop_2010,Katz2011, akbulut2011focusing} to subwavelenght microscopy.\cite{PhysRevLett.106.193905} Merging nonlinear optics with adaptive techniques is much more difficult for the inherent instability of nonlinear processes. However one-dimensional numerical simulations\cite{PhysRevLett.109.033903} predict that such an approach is possible for Random Lasers (RLs).

RLs\cite{Wiersma08} are coherent light sources in which stimulated emission is generated in a "disordered" cavity, with a certain degree of localization,\cite{Fallert_coexistence_nature, Lagendijk_2010_modestructure,Cao_Confinement} infiltrated with a gain material. Here \emph{disordered} means that the cavity has not been previously designed but is randomly selected by the light diffusion process inside the random medium.\cite{PhysRevE.54.4256} Therefore the position, direction and wavelength at which the lasing action occurs are unpredictable \emph{ex-ante}. From the fundamental point of view RLs are paradigmatic systems that mix intriguing features like nonlinearity, disorder and complexity, but are becoming increasingly more attractive also from the applied point of view.\cite{redding2012speckle,cerdan2012fret}

In trying to control the RL emission spectrum various approaches have been proposed to date such as the design of individual scattering elements\cite{gottardo2008resonance} or the engineering of the absorption.\cite{el2011tuning} In this paper we demonstrate another strategy that allows an active, sub-nanometer precise control over the spectrum emitted by the RL. The procedure relies on the selective excitation of modes through pumping engineering. The hypothesis behind this work is that pumping configurations can be found that couple best to a (preselected) target mode so that its emission is enhanced whereas other modes' is suppressed. To demonstrate the effect a single, isolated, small cluster (average diameter between 10 and 25 micrometer, 10 micrometers in thickness) of TiO$_2$ (sub-micrometer) nanoparticles embedded in a laser dye solution is the best candidate.\cite{PhysRevA.85.043841, leonetti:051104, Leonetti2011} At variance with small clusters, larger assemblies present an enormous amount of modes,\cite{lawandy1994laser} all practically indistinguishable, in which the selection of a single one is all but impossible.

In order to realize a pumping scheme that optimizes the lasing action from a single mode we exploited the same approach of ref \cite{PhysRevA.85.043841} in which the amplified spontaneous emission (ASE) of the dye surrounding a scattering cluster is used for pumping. In this configuration the cluster lies on the surface of a microscopy coverslip and is embedded in the rhodamine solution, which is pumped from the bottom while the emitted radiation is collected from the top.

The numerous laser modes available in the cluster have different coupling abilities with light coming from a given direction and impinging on a given spot of the cluster. Ideally it would be desirable to be able to send a pump laser beam with any direction and pointed at any desired spot on the cluster. This can be achieved indirectly by generating ``rays'' of ASE in the surrounding dye with a "green" laser in a manner entirely similar to that used to determine gain in the stripe length method.\cite{sapienza2011optical} A green laser excites "red" ASE from the dye and these red beams pump the cluster. The cross section of the green laser beam determines the  direction in which ASE is generated. If additionally a circular area encompassing the whole cluster is illuminated, preparing the RL cluster near lasing threshold, the rays will trigger lasing only for modes that couple best with the incoming ASE. This approach can be called directionally constrained pumping. At variance with the spatial constraint a directional constraint allows to pump over the threshold modes that are located in distant positions within the cluster thus being barely interacting.

\begin{figure}
\includegraphics[width=8 cm]{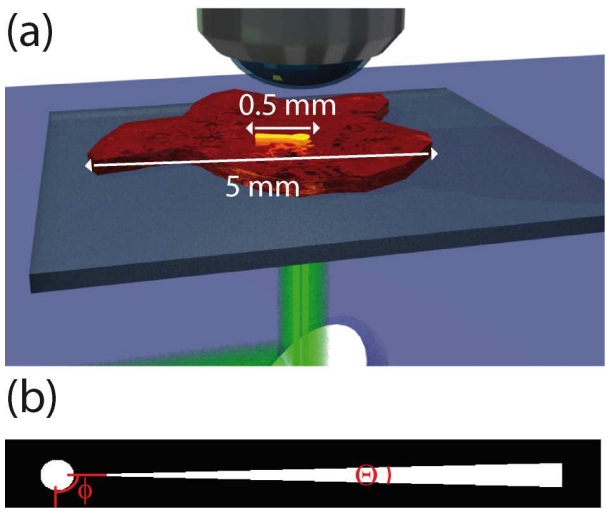}
\caption{(Color Online)a) A sketch of the setup for tuning a random laser. Light, generated by a Nd:YAG pulsed laser (9 ns, 532 nm, 20 mJ max pulse energy) is modulated by the SLM and then directed onto the sample from underneath. Average dimension of the rhodamine drop and of the wedge are indicated in the figure. The RL emission is collected from a microscope objective (0.55 numerical aperture) shown in the upper part of the figure.  b) A generic mask sent to the SLM with the parameters $\theta$ and $\phi$ indicated.
\label{setup}}
\end{figure}

In a nutshell, the goal is to optimize the pumping direction to favor emission from a target mode chosen among the many that may be activated in the RL cluster. To achieve such a goal, we used a spatial light modulator (SLM) in amplitude configuration (a typical gray scale bit-map sent to the SLM is shown in panel \ref{setup}b) that shapes the green laser (see reference \cite{PhysRevA.85.043841}) and subsequently the ASE beams (yellow area in Fig. \ref{setup}a).

Since the ASE is directional a circular sector (that we call wedge) pumped by the green laser generates an intense ASE beam at the vertex and is the best configuration to control both the angular span of the pumping and the exact point in which the ASE is impinging on the cluster. The wedge is pointing to the center of a circular area which maintains the modes of the cluster barely below threshold while the wedge feeds the modes that have to be brought to lasing, actively performing the directional selection. In previous experiments multiple wedges have been used to guarantee the hydrodynamic stability of the solution. In the present experiment the angular span of the wedge (and the total pumped area) is always small preventing the presence of hydrodynamic fluxes in the liquid. This technique allows a completely digital control over the green pump beam cross section, hence the shape of the population inverted area on the sample and, subsequently, over the directions ASE generated. Instead of using a stripe we used a pointed wedge (a circular sector) whose angular span ($\theta$ parameter) and orientation ($\phi$ parameter see figure \ref{setup}b) can be selected. The pumping conditions can therefore be modified without any mechanical intervention on the experimental setup.

\begin{figure}
\includegraphics[width=8 cm]{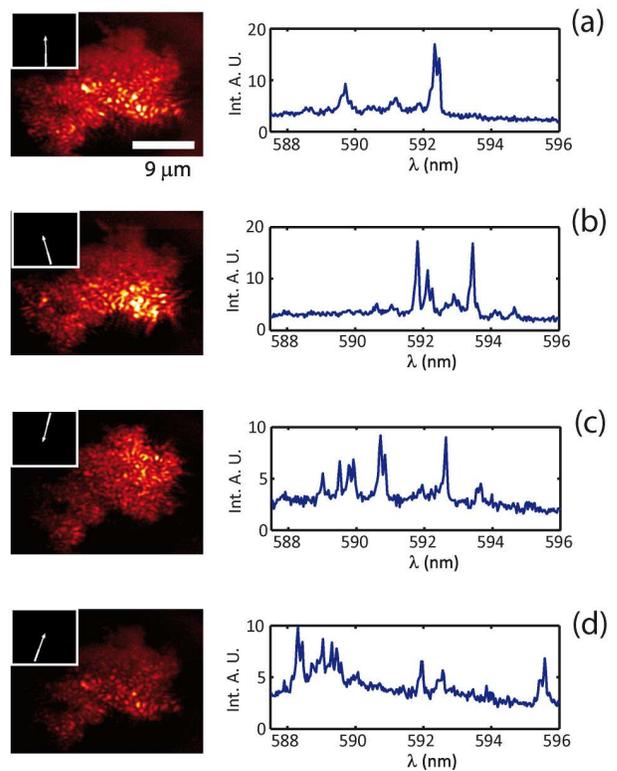}
\caption{(Color Online) Images of cluster C1 for 4 different pumping configurations (in the insets) and relative spectra. Scale bar is 9 $\mu$m \label{phiscan}}
\end{figure}

The initial process in the optimization protocol involves identifying the available modes which can be achieved by retrieving the RL emission as the input direction $\phi$ is scanned. In figure \ref{phiscan} a-d we present images of an individual cluster (sample C1) pumped under 4 arbitrary values of $\phi$ (indicated in the respective upper left insets). As can be seen, ASE pump rays with different input directions result in modes (hotsposts in the cluster images) dwelling in different locations within the cluster. Light emitted from the cluster can also be spectrally analyzed. For this purpose a microscope objective images the cluster on the tip of a fiber that relays the signal to a spectrograph that provides the spectrum corresponding to light emitted in a circular area 20 micrometers in diameter centered on the cluster. Spectra corresponding to each configuration are shown on the right panels of figures \ref{phiscan}a-d.

Now we want to find a configuration that allows the selective pumping of an individual mode while suppressing all other contributions. Since the response matrix of the selected cluster is not known we proceed by a sequence of steps. First, we select the wavelength for which we want to maximize the emission (the "target frequency"). Then we try different pumping configurations (input direction) from which we select that showing strongest signal at the target frequency. And finally we change pumping parameters to suppress unwanted modes (other peaks in the spectrum). The optimization protocol may be performed manually or by a computer programmed routine. In the following we will refer to the configuration obtained at the end of an optimization stage as the \emph{best configuration} (BC). Although the system is static (the cluster is fixed and spectra are composed of the same peaks shot after shot) successive pulses have slight intensity fluctuations,\cite{Leonetti2011} in order to get rid of which, spectra are averaged over 25 shots.
The optimization then proceeds through the following stages. The \textbf{first step} is a coarse scan over $\phi$: we change the direction of the input ASE ray by changing the orientation of the pumping wedge. The scan is performed with an angular resolution of 1$^\circ$, fixed $\theta=6^\circ$ and fixed fluence over the sample of 0.1 nJ/$\mu m^2$ and allows to find the BC. The \textbf{second step} is a fine scan, performed with angular resolution (0.1 $^\circ$) around the value of $\phi$ retrieved from the previous optimization (within a range of $\pm$2$^\circ$ ). Again the BC is that with the highest intensity at the target frequency. This configuration, referred to as the \textit{preliminary optimization}, is aimed at coupling the maximum amount of energy onto the target mode.

\begin{table}[h]
\begin{tabular}{c c c c c }
\hline \hline
Step  &  Scan. Var.     &   Range                       &  Step               &   Max. Magn.                  \\ \hline
 1    &  $\phi$         &   360  $^\circ$               &  0.5$^\circ$        &   $I_{target}$                \\
 2    &  $\phi$         &   $\pm$ 2$^\circ$ around BC   &  0.1$^\circ$        &   $I_{target}$                \\
 3    &  $\theta$       &   0.2$^\circ$-6$^\circ$       &  0.1$^\circ$        &   $SNR$                       \\
 4    &  $\phi$         &   $\pm$ 2$^\circ$ around BC   &  0.1$^\circ$        &   $SNR$                       \\
 5    &  Fluence        &   0.05-0.20 nJ/$\mu m^2$      &  0.01 nJ/$\mu m^2$  &   $SNR$                     \\ \hline \hline
\end{tabular}
\caption{\label{tab:resume} Parameters varied in each stage of optimization, range and step of variation and magnitude optimized. Scan. Var. stands for ``scanned variable'', while Max. Magn. stands for ``maximized magnitude''. }
\end{table}

At this stage we have strongly increased the intensity at the target mode, but the intensity from other peaks may still be considerable. Thus the following steps are aimed at maximizing the signal-to-noise ratio defined as $SNR = I_{target}/I_{OM}$  where $I_{target}$ is the intensity at the target mode while $I_{OM}$ is the maximum of the other modes' intensities. An inverse approach (\emph{i.e.} maximizing first the $SNR$ and then the Intensity), is more difficult because, before the intensity optimization, the identification of a single peak in a messy spectrum is challenging. In our protocol the \textbf{third step} is an optimization over the $\theta$ parameter in search for a compromise between increasing the energy reaching the target mode without exciting others. Here $\theta$ is scanned from 0.2$^\circ$ to 6$^\circ$ with a step of 0.2 $^\circ$. The chosen configuration is the one for which $SNR$ is maximum. In the \textbf{fourth step} we repeat a fine tuning over $\phi$ performed with the newly defined $\theta$ (scanning parameters are the same used in the second step). Again the BC is the one that maximizes the $SNR$. The \textbf{fifth step} is a scan over the input fluence that is varied between 0.05 and 0.2  nJ/$\mu m^2$ with steps of 0.01  nJ/$\mu m^2$. As previously the chosen configuration is the one with the highest $SNR$. A summary of all the steps is presented in table \ref{tab:resume}.

\begin{figure}
\includegraphics[width=8 cm]{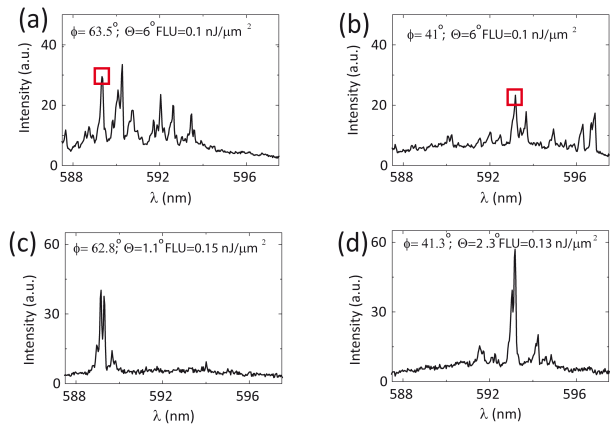}
\caption{(Color Online)  Spectra obtained for two different target frequencies at the end of the preliminary optimization (a and b) and at the end of the optimization process (c and d). The experimental parameters are indicated in each panel (FLU stands for Fluence).
\label{Optimized}}
\end{figure}

In figure \ref{Optimized} we show the spectra obtained after the preliminary optimization (end of the second optimization step, panels a-b) and at the end of the complete optimization protocol (after the fifth optimization step, panels c-d). The striking feature of this results is that modes other than the target are practically suppressed: $SNR$ is 4.3 in panel c and 3.4 in panel d. Thus in the preliminary optimization the intensity is coupled strongly to the mode of interest while in the three final steps the pumping is sculpted to lower the contribution from unwanted modes.

The stability of the system is an important point. The system is a static one in which gain (dye) and scattering material (nanoparticles) arrangements are fixed and modes are persistent so that the system, at variance with for instance colloidal ones, stays the same throughout the experiment. This was plainly shown by speckle patterns taken before and after sequences of experiments \cite{PhysRevA.85.043841}. We verified that in a time comparable with the duration of the protocol (about 600 of seconds) just one mode over twenty is lost. Spectrum intensity fluctuation in our system are comparable with that reported in \cite{Leonetti2011} so that averaging over 25 shots enables the optimization procedure.

This work only aims at providing a proof of principle and demonstrate that it is possible to select modes in a RL and although the selection protocol is hardly optimized it clearly shows its potential. In fact, the presented optimization protocol allows appreciable results for all the investigated wavelengths. We will next demonstrate the possibility to use the device to obtain narrow radiation emission at any user selected frequency chosen across the whole active spectral range of the RL (in the region in which RL peaks are found).

\begin{figure}
\includegraphics[width=8 cm]{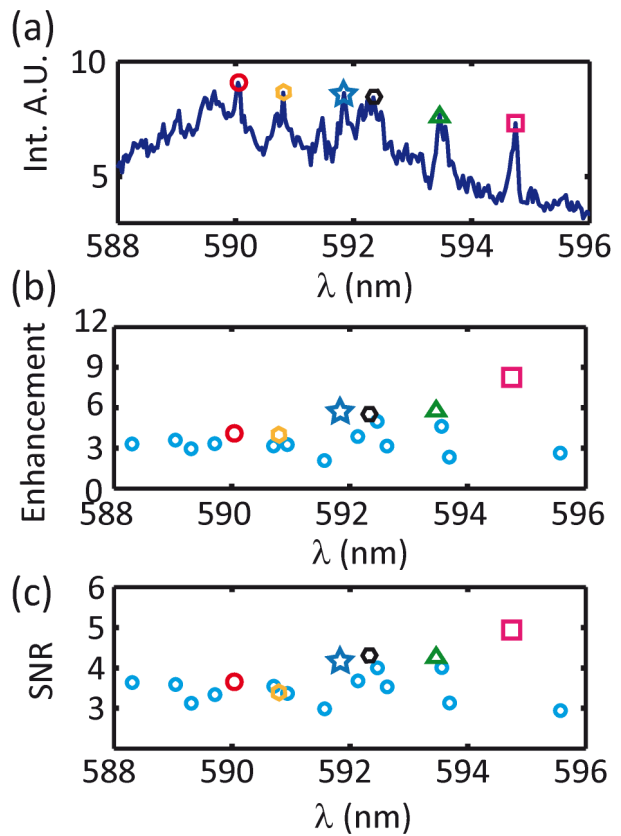}
\caption{(Color Online)   a) Average spectrum $\langle S \rangle_{\phi}$  obtained by averaging the spectra measured at the first optimization step (configurations with different values of $\phi$ ($\theta$ fixed to $6^\circ$)). In panel b) the enhancement $I_{OPT}/I_{AVG}$ as a function of the wavelength. In panel c) the  $SNR$ is reported instead.
\label{magnification}}
\end{figure}

To measure the spectral effectiveness of our protocol, we will now compare the optimized spectrum for any given mode with the average spectrum, the most unfavorable standard because it includes modes showing under all possible pumping conditions. Thus to have a measure of the average response of the cluster we measured $\langle S (\lambda)\rangle _{\phi}$, that is the spectrum resulting by averaging spectra over $\phi$. In practice $\langle S(\lambda) \rangle _{\phi}$ measures the response of the cluster regardless of pumping configuration. Figure \ref{magnification}a shows $\langle S (\lambda) \rangle_{\phi}$ obtained by averaging over 360 equally spaced values of $\phi$.

To characterize the spectral improvement upon optimization figure \ref{magnification}b shows the enhancement defined as the ratio of the intensity at target frequency at the end of the optimization process $I_{OPT}$ to the value of the intensity at the same frequency in the average spectrum $I_{AVG}$ for nineteen different target frequencies. The values retrieved are higher than 1 meaning that every mode can be optimized with respect to its value in an average configuration. A few outstanding modes in the average spectrum, marked by different symbols, show remarkably good performances when optimized.

Other important parameters are the $SNR$ (presented in figure \ref{magnification}c) and the average linewidth that is 0.3 nm. Being both $SNR$ and the enhancement higher than 3 for most of the investigated frequencies means that it is practically possible to obtain a line narrowed tunable emission throughout the active spectral range of the RL.

\textbf{}

{\it Conclusions ---}
We demonstrated a random laser capable of generating a tunable (within a fraction of nm) lasing emission at any pre-selected wavelength throughout the lasing spectrum. The monochromatic emission tunability has been obtained by a feedback mechanism that finds the configuration which allows to pump an individual mode among the many that are available for activation in the disordered matrix of the cluster, and suppressing the contribution from competing modes. This is the tunable dye-based source with the smallest resonator producing a subnanometer emission ever developed. Improvement of the rudimentary optimization protocol used here will reduce the time and increase the mode selectivity improving both $SNR$ and enhancement. Additional challenges comprise the optimization of the quality of the lasing material, (employing polymeric or solid active media), which will allow to increase the stability and efficiency of the device.

\begin{acknowledgments}
The research leading to these results has received funding from the EU FP7 NoE Nanophotonics4Enery Grant No 248855; the Spanish MICINN CSD2007-0046 (Nanolight.es); MAT2012-31659 (SAMAP) and Comunidad de Madrid S2009/MAT-1756 (PHAMA).
\end{acknowledgments}

%

\end{document}